# High-Capacity Metasurface at Limits of Polarization and Wavelength Multiplexing

Yanjun Bao[1]\*, Hongsheng Shi[1], Rui Wei[1], Boyou Wang[1], Zhou Zhou[2], Cheng-Wei Qiu[2]\*

and Baojun Li[1]\*

[1]Guangdong Provincial Key Laboratory of Nanophotonic Manipulation, Institute of Nanophotonics, College of Physics & Optoelectronic Engineering, Jinan University, Guangzhou 511443, China

[2]Department of Electrical and Computer Engineering, National University of Singapore, 4 Engineering Drive 3, Singapore 117583

\*Corresponding Authors: Y. Bao (yanjunbao@jnu.edu.cn), C. Q. (chengwei.qiu@nus.edu.sg), B. Li (baojunli@jnu.edu.cn)

**Abstract:** Polarization and wavelength multiplexing are the two most widely employed techniques to improve the capacity in the metasurfaces. Existing works have pushed each technique to its individual limits. For example, the polarization multiplexing channels working at a single wavelength have been significantly increased by using noise engineering. However, it is still challenging to achieve the multiplexing limits of wavelength and polarization simultaneously. Besides, such multiplexing methods suffer from computational inefficiencies, hindering their application in tasks like image recognition that require extensive training computation. In this work, we introduce a gradient-based optimization algorithm using deep neural network (DNN) to achieve the limits of both polarization and wavelength multiplexing with high computational efficiency. We experimentally demonstrate this capability, achieving a record-breaking capacity of 15 holographic images across five wavelengths and the maximum of three independent polarization channels, as well as 18 holographic images across three wavelengths and six corelated polarization channels. Moreover, leveraging the high computational efficiency of our DNN-based method, which is well-suited for processing large datasets, we implement large-scale image recognition tasks across 36 classes encoded in a record of nine multiplexed channels (three wavelengths × three polarizations), achieving 96% classification accuracy in calculations and 91.5% in experiments. This work sets a new benchmark for high-capacity multiplexing with metasurfaces and demonstrates the power of gradient-based inverse design for realizing multi-functional optical elements.

# Main

Metasurfaces are ultra-thin, planar structures that manipulate electromagnetic waves in ways that traditional materials cannot[1-3]. These engineered surfaces can precisely control the phase, amplitude, and polarization of light, facilitating a range of applications in lensing[4-8] and holography[9-15]. Recently, multifunctional metasurfaces have attracted significant attention for their ability to perform multiple independent optical operations within the same structure, thereby reducing device size and complexity while enhancing performance. To achieve these capabilities, multiplexing methods[16-21] that enable different optical properties to coexist and function independently are typically employed. Polarization and wavelength multiplexing are the two most widely employed techniques, which enable the independent manipulation of light based on various polarization states or wavelengths.

Multiplexing limits for polarization and wavelength refer to fully utilizing the maximal number of channels achievable for each. For polarization, the planer metasurface structure has a maximal channel number of three[22] due to the mirror symmetry of the structure, which has be realized with composite pixel designs[23, 24]. Recently, the engineered noise is introduced to break this limitation with 11 polarization channels[25], but the independent channel number remains three. Wavelength, being a continuous parameter, theoretically offers infinite channels. However, practical constraints such as finite bandwidth response limitations and crosstalk typically restrict wavelength multiplexing to a finite number ($N$) of channels. Consequently, multiplexing limits of polarization and wavelength is characterized by a total multiplexing channel number of 3$N$ (as illustrated in Fig. 1a), representing the product of the three polarization channels and $N$ wavelength channels.

On the other hand, computational efficiency is crucial for the practical application of metasurface design algorithms, particularly in tasks like image recognition[26] that involves large-scale data training process. For wavelength multiplexing, it usually presents a particular challenge in forward design due to the difficulty of finding geometric parameters that satisfy independent optical responses across multiple wavelengths. One solution is to increase the geometric degrees of freedom by employing composite pixels and sweeping various geometries to find the best ones to match the design targets. For instance, a composite pixel with two rectangle strips can focus at the same point across three wavelengths[27]. A composite pixel containing two nanorods was utilized to achieve an achromatic metalens over a broadband spectrum[28]. These forward designs have a multiplexing number of 1$N$ and require extensive geometric sweeps, resulting in slow computation (Fig. 1a). To increase multiplexing from 1$N$ to 2$N$, other algorithm, such as genetic algorithms (GA)[29], statistic machine-learning-based inverse design (ID)[30] have been proposed. However, they also require numerous random cases in each iteration or large datasets for training, making them computationally expensive and slow to converge. In contrast, gradient-based optimization algorithms leverage gradient information to guide the search process, allowing for rapid convergence. These algorithms can significantly enhance computational efficiency as they can decay rapidly and only require a few cases in each iteration. For example, with the adjoint method[31], each iteration only calculates twice to obtain the gradient information, thus enhancing computational efficiency. Despite these advancements, previous approaches have not achieved multiplexing limits (3$N$), and their algorithms still lack optimal computational efficiency, hindering broader applications in metasurface design.

In this work, we introduce a gradient-based optimization algorithm compatible with graphics processing unit (GPU) acceleration of batch gradient calculation, achieving 3$N$ multiplexing limits of polarization and wavelength (Fig. 1a, red pentacle). This method sets a new benchmark for multiplexing channels, with the upper limit of three independent polarization channels and at least five wavelength channels in our measurement. With this approach, we further demonstrate six "limit-breaking" polarization channels with three wavelength multiplexing, significantly boosting multiplexing capacity. Moreover, our algorithm incorporates a deep neural network (DNN) compatible with GPU acceleration, making it highly effective and computationally efficient for processing and training large datasets. To demonstrate its capabilities, we employ the metasurface for an image recognition task, simultaneously operating on a record number of nine independent polarization-wavelength multiplexed channels. Our work achieves multiplexing limits with high computational efficiency for extensive calculations, potentially inspiring new multifunctional optical element designs.

**Design principle**

As a primary demonstration of the multiplexing limits, we encode independent holographic images multiplexed across three polarization channels ($J_{11}$, $J_{12}$, and $J_{22}$) and $N$ wavelength channels ($\lambda_1$, $\lambda_2$, ..., $\lambda_N$), resulting in $3N$ independent images, as shown in Fig. 1b. For the metasurface design, we select the nanoblock structure as the fundamental unit, characterized by three geometric parameters: length ($W_1$), width ($W_2$), and rotational angle ($\theta$). While more intricate unit designs may offer additional geometric degrees of freedom, nanoblock structures represent the most viable option within the constraints of fabrication limitations at hundreds of nanometers periods. The nanoblock can be described by the Jones matrix of a linearly birefringent wave plate:

$$J(\lambda_i) = R(-\theta)\begin{bmatrix} A(\lambda_i) & 0 \\ 0 & B(\lambda_i) \end{bmatrix} R(\theta) \qquad (1)$$

where $A$ and $B$ represent the complex-amplitude coefficients of light linearly polarized along the fast and slow axes, respectively, both of which are wavelength-dependent. The obtained Jones matrix is a symmetric 2×2 matrix with three independent components: $J_{11}$, $J_{22}$, and $J_{12}/J_{21}$.

In conventional single-wavelength metasurface hologram design, the typical approach involves calculating the metasurface phase distribution and subsequently determining the geometric size to match the desired phases. However, this approach is unsuitable for multiwavelength multiplexing, as the coefficients $A$ and $B$ are correlated in wavelength dimension, making it challenging to find a suitable geometric size to match the target phase across all wavelengths.

Fig. 2a illustrates the optimization algorithm employed to achieve wavelength and polarization multiplexing, where the three parameters of nanoblocks are selected as starting points of the algorithm as they remain unchanged across various wavelengths, unlike coefficients $A$ and $B$. Gradient information is crucial in the optimization algorithm, facilitating faster convergence. To obtain gradient information regarding the optical coefficients with respect to the geometric parameters at various wavelengths, we conduct numerical calculations and sweep $A$ across different $W_1$ and $W_2$ values under diverse wavelengths. These data subsequently serve as training inputs for a DNN (Fig. 2b) to establish the relationships. The DNN has two input parameters ($W_1$ and $W_2$) and $2N$ output neurons, where $N$ is the wavelength number, with each wavelength represented by two neurons corresponding to the real and imaginary parts of $A$. The case for $B$ can be obtained by using the same DNN by swapping $W_1$ and $W_2$ values due to symmetry. Given that DNN inherently provide gradient information, we can derive the gradients of $A$ and $B$ concerning $W_1$ and $W_2$ values. This DNN architecture is compatible with GPU acceleration and offers advantages for large-scale data training, as demonstrated subsequently. Further details regarding this DNN are provided in the Methods and Fig. S1 of Supplementary Material.

Utilizing the obtained coefficients $A$ and $B$, we calculate the three components of the Jones matrix ($J_{11}$, $J_{12}$, and $J_{22}$) across different wavelengths according to Eq. 1. Subsequently, these Jones matrix components are used to compute the holographic images [$H_1(\lambda_i)$, $H_2(\lambda_i)$ and $H_3(\lambda_i)$] using diffraction propagation theory. The mean squared error between the calculated holographic images $H_j(\lambda_i)$ and the designed targets $H'_j(\lambda_i)$ are served as the loss function. Given that the entire process is gradient-calculable, the geometric parameters of the nanoblock ($W_1$, $W_2$ and $\theta$) can be optimized to minimize the loss and ultimately realize the desired functionalities.

As a demonstration, we choose $N=5$, with five wavelengths: 532 nm, 650 nm, 715 nm, 808 nm, and 860 nm. It is important to note that this algorithm is not limited to five wavelengths but can be applied for a larger number (e.g., $N=9$ in Fig. S2 of Supplementary Material). Figure 2c illustrates the loss versus iteration number, demonstrating stable decay and convergence after approximately 400 iterations. The resulting 15 holographic images (Fig. 2d) align closely with our designed images. While minor background stray light may be present, all images remain distinct from one another, representing the largest channel number of polarization and wavelength multiplexing known to date.

To experimentally validate this concept, we fabricated the crystal silicon metasurface with a height of 600 nm and a period of 250 nm (Fig. 3a) using the standard electron beam lithography process (see details in Method section). The nanoblock metasurface period is well below half the smallest wavelength (532 nm), ensuring effective diffraction. The experimental setup is shown in Fig. S3 in Supplementary Material. A continuous spectrum laser source is used with specific wavelengths isolated by an acousto-optic tunable filter. The incident light was polarized by a linear polarizer before being incident upon the metasurface. The holographic images were collected by an objective, filtered by another linear polarizer, and finally imaged on the CMOS sensor with a lens. The measured intensity profiles with various wavelengths and polarizations (Fig. 3b) closely match the design and calculated images, capturing significant details. Minor discrepancies between the experimental and calculated results may be attributed to fabrication imperfections and overlooked coupling between unit periods.

The efficiency at various wavelengths is also measured and compared to the calculated results (Fig. 3c). Here, efficiency is defined as the average efficiency of $x$ and $y$ incident polarization incidences, given by the equation: $\eta_{12}+(\eta_{11}+\eta_{22})/2$, where $\eta_{ij}$ represents the efficiency of polarization channel $J_{ij}$. The maximal measured efficiency reached 25% (33% in calculation) at a wavelength of 808 nm. Due to the inherent increased optical absorption of silicon materials at higher frequencies, the efficiency dropped to 13% (17% in calculation) as the wavelength decreases to 532 nm.

**Wavelength and "limit-breaking" polarization multiplexing**

From a physical standpoint, the upper limit of independent polarization channels is three for single-layer structure. Recently, engineered noise has been introduced to surpass this limitation[25], albeit at the expense of reducing the imaging quality of each channel. However, the number of independent channels remains capped at three and operates under a single wavelength. Here, we extend this "limit-breaking" polarization multiplexing to include wavelength multiplexing (Fig. 4a), which naturally fits within the framework of our proposed approach.

We select six different combinations of input $|\alpha_i\beta_i\rangle$ and output $|\alpha_o\beta_o\rangle$ polarizations at three different wavelengths (532 nm, 650 nm, and 808 nm). The output complex fields at metasurface surface can be expressed as:

$$S(\lambda_i) = \langle \alpha_o\beta_o | J(\lambda_i) | \alpha_i\beta_i \rangle \tag{2}$$

where $|\alpha_i\beta_i\rangle = [\cos\alpha_i, \sin\alpha_i e^{i\beta_i}]^T$ and $\langle\alpha_o\beta_o| = [\cos\alpha_o, \sin\alpha_o e^{-i\beta_o}]$. Then the holographic images are calculated, and the optimization algorithm follows the same procedure as outlined in Fig. 2a. Given that there are only three independent polarization channels, correlations among the six channels are predictable, leading to inevitable cross-talk. As depicted in the

calculated results (first column for each wavelength in Fig. 4b), all images prominently feature the main subject with maximum intensity, while cross-talk appears as weaker intensity in the background. The measured intensity profiles (second column for each wavelength in Fig. 4b) closely align with the calculation, and their independence is clearly observable. Notably, we observe lower cross-talk in the experiment than in the calculation. We attribute this discrepancy to noise introduced by fabrication imperfections and the neglect of coupling, both of which contribute to reducing the sharpness of background cross-talk.

**Image recognition with nine channels**

Lastly, we highlight the advantages of our design method with high computational efficiency, which is very suitable for large-scale data computation. Metasurfaces have recently been utilized for image recognition, but predominantly with a single wavelength and a few polarization channels (one or two)[32, 33]. For image recognition task, it necessitates large-scale data training, requiring high computational efficiency and easily gradient calculations. Our proposed gradient-based optimization algorithm based on DNN is highly compatible with modern GPU-based batch gradient calculation, rendering it particularly suitable for this task within short time.

To illustrate, we select three wavelengths and three polarization channels, resulting in a total of nine channels. In each channel, four classes of images are classified, yielding a total of 36 classes (from 0 to 9 and A to Z) (Fig. 5a). Within each channel, the holographic image plane is divided into four discrete regions representing the classification of the four labels, respectively. The loss function considers all nine channels and aims to maximize the optical signals in the corresponding detection regions. Further details about multi-channel image recognition training with metasurfaces can be found in the Methods section. Figure 5b illustrates the loss and recognition accuracy versus iteration for both training and validation datasets, which converge after approximately 30 epochs. Both training and validation losses decrease over iterations and converge to stable, low values, demonstrating the effectiveness of the training process. The accuracy of the validation datasets approaches 96% for the best values (almost the same as the training datasets), which is high for image recognition tasks, especially considering the use of only a single layer of metasurface. We also verify the accuracy of cases with fewer channel numbers, achieving 99.6% accuracy for one channel and 98.8% for three channels (Fig. 5c). Although the accuracy of image recognition decreases with an increase in channels, the decrease is marginal, maintaining a 96% accuracy for nine channels.

For experimental validation, we randomly select 40 images per classification category from the validation datasets (204 images per classification). The results are presented using the confusion matrix (Fig. 5e), with the measured accuracy being 91.5% according to the experimental statistical results. The calculation confusion matrix is also presented in Fig. 5d for comparison. In the experiment, input images with varied intensity are generated using phase-only modulation[34, 35] based on a spatial light modulator (SLM) . The measured input images closely match the designed input label images (Fig. 5f). The optical intensity distributions at the holographic image plane obtained in calculation and experiment are shown in Fig. 5g-h, with the maximum energy accurately accumulated in the target region (Fig. 5i). Here, we only present selected three channels. The details of the full nine channels can be found in Fig. S4 in in Supplementary Material. The results presented herein demonstrate the feasibility of our design for image recognition with significantly larger number of channels.

**Conclusions**

In summary, we have introduced a gradient-based optimization algorithm leveraging DNN to significantly enhance the performance of polarization and wavelength multiplexing with metasurfaces. Our approach has two main advantages: achieving the multiplexing limits and high computational efficiency. With these benefits, we experimentally demonstrated the capacity-limit multiplexing with a record-breaking of 15 independent holographic images and extended the "limit-breaking" polarization multiplexing to include three wavelength multiplexing, significantly boosting the multiplexing capacity. The high computational efficiency of our algorithm makes it suitable for large-scale image recognition tasks. Using a single-layer metasurface, we achieved 96% classification accuracy in calculations and 91.5% in experiments across 36 classes encoded in nine multiplexed channels. Our work sets a new benchmark for high-capacity multiplexing with metasurfaces and demonstrates the potential for realizing multifunctional optical elements. We anticipate this approach will inspire new classes of compact optical systems for applications such as imaging, sensing, and optical information processing.

**Methods**

**Sample fabrication.** The metasurface sample is fabricated with the standard electron beam lithography process. Initially, a commercial silicon-on-insulator (SOI) wafer with a 1200 nm thick device layer is bonded to a glass substrate via adhesive wafer bonding and deep reactive ion etching (DRIE) techniques.

Next, the device layer's thickness is reduced to 600 nm using inductively coupled plasma (ICP) etching. Afterward, a 300 nm-thick layer of hydrogen silsesquioxane (HSQ) is spin-coated onto the substrate at 4000 rpm and baked on a hot plate at 90 °C for 5 minutes. Subsequently, a 30 nm-thick aluminum layer is deposited on top of the HSQ layer by thermal evaporation to function as a charge dissipation layer. The metasurface pattern is then written on the HSQ layer using electron beam lithography with an acceleration voltage of 30 kV. Following exposure, the aluminum layer is removed with a 5% phosphoric acid solution, and the resist is developed in tetramethylammonium hydroxide (TMAH) for 2 minutes at room temperature. Inductively coupled plasma-reactive ion etching (ICP-RIE) is then used to transfer the pattern into the silicon film. Finally, the samples are immersed in a 10% hydrofluoric (HF) acid solution for 15 seconds to remove any remaining HSQ mask, rinsed with deionized water, and dried with nitrogen gas.

**DNN for establishing the relationship between optical coefficients and geometric parameters.** We used a fully connected linear DNN with 6 layers, each containing 200 neurons, to establish the relationship between optical coefficients and geometric parameters. The input layer has two neurons ($W_1$ and $W_2$), and the output layer has $2 \times N$ neurons, corresponding to the real and imaginary parts of optical coefficient $A$ at $N$ wavelengths. A rectified linear unit (ReLU) activation function was used between each layer. For training data, we first used Finite-Difference Time-Domain (FDTD) simulations to obtain the optical coefficient $A$ across 36 linearly spaced values (from 60 nm to 200 nm) for $W_1$ and $W_2$, resulting in 1296 datasets. Optical coefficients $B$ were derived by swapping the mapping of $W_1$ and $W_2$ due to symmetry. Subsequently, we interpolated the 1296 data points to expand the dataset size to 20,000. The network was trained using a batch size of 5000 and an Adam optimizer with a learning rate set at 0.01. The dataset was divided into train and validation sets (80% and 20%, respectively). The training loss was used to generate the gradients, and training was stopped when the validation loss stopped improving. Fig. S1 and Fig. S2 illustrate the train and validation loss, along with the comparison between predicted and true values for the 5-wavelength and 9-wavelength scenarios, respectively.

**Image recognition algorithms.** The image recognition dataset was selected from the Chars74k dataset, with each label containing 1016 images. We focused on the first 36 labels (from 0 to 9 and A to Z) for training image recognition across 9 channels, with each channel classifying 4 labels. For each label, 80% (812) of the images were allocated for training, leaving 20% (204) for validation, resulting in a total of 29,232 training images and 7,344 validation images. The loss function plays a pivotal role in determining the final classification accuracy and efficiency within designated regions. We defined the loss function as: Nllloss[Log($P_d$)] + 0.4×CrossEntropyLoss($P_d$). Here, Log denotes the base-10 logarithm, Nllloss and CrossEntropyLoss are the standard loss functions in Pytorch, and $P_d$ represents the optical power at the designated region. This definition effectively balances efficiency and classification accuracy. The network was trained using a batch size of 32 and an Adam optimizer with a learning rate set to 0.001.

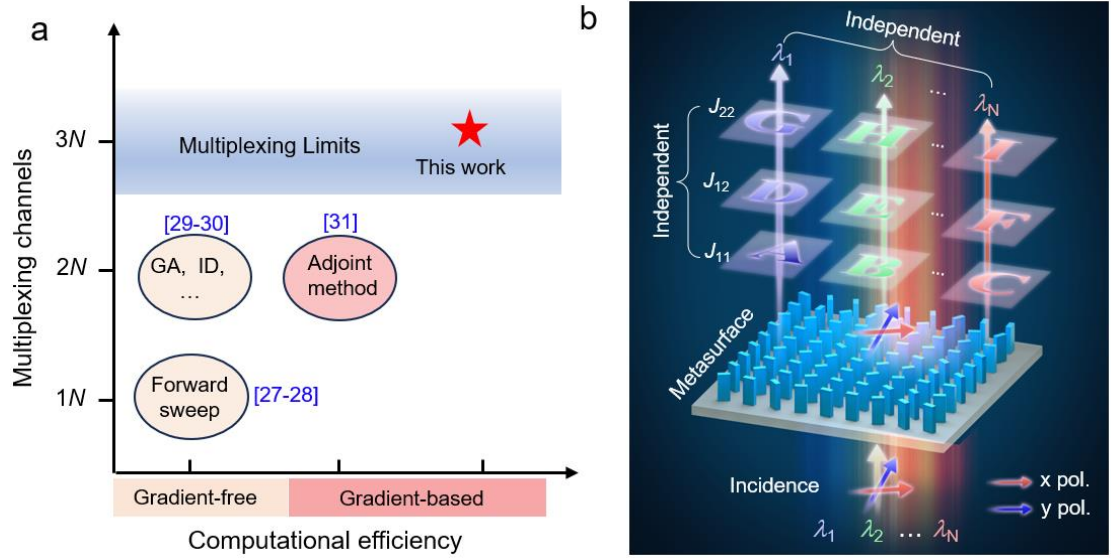

**Fig. 1. Realization of multiplexing limits of polarization and wavelength with high computational efficiency. a**, A summary of multiplexing channels and computational efficiency for polarization and wavelength multiplexing using metasurfaces in the literature. The multiplexing number for wavelength is denoted as $N$, and the maximum number for polarization is three, resulting in a multiplexing limit number of $3N$. The red pentacle indicates our work, which realizes multiplexing limits with the highest computational efficiency. **b**, A schematic view of holographic demonstration for multiplexing limits of polarization and wavelength. The metasurface enables three polarization ($J_{11}$, $J_{12}$, and $J_{22}$) and $N$ wavelength channels ($\lambda_1$, $\lambda_2$, ..., $\lambda_N$), with all generated holographic images being independent with each other.

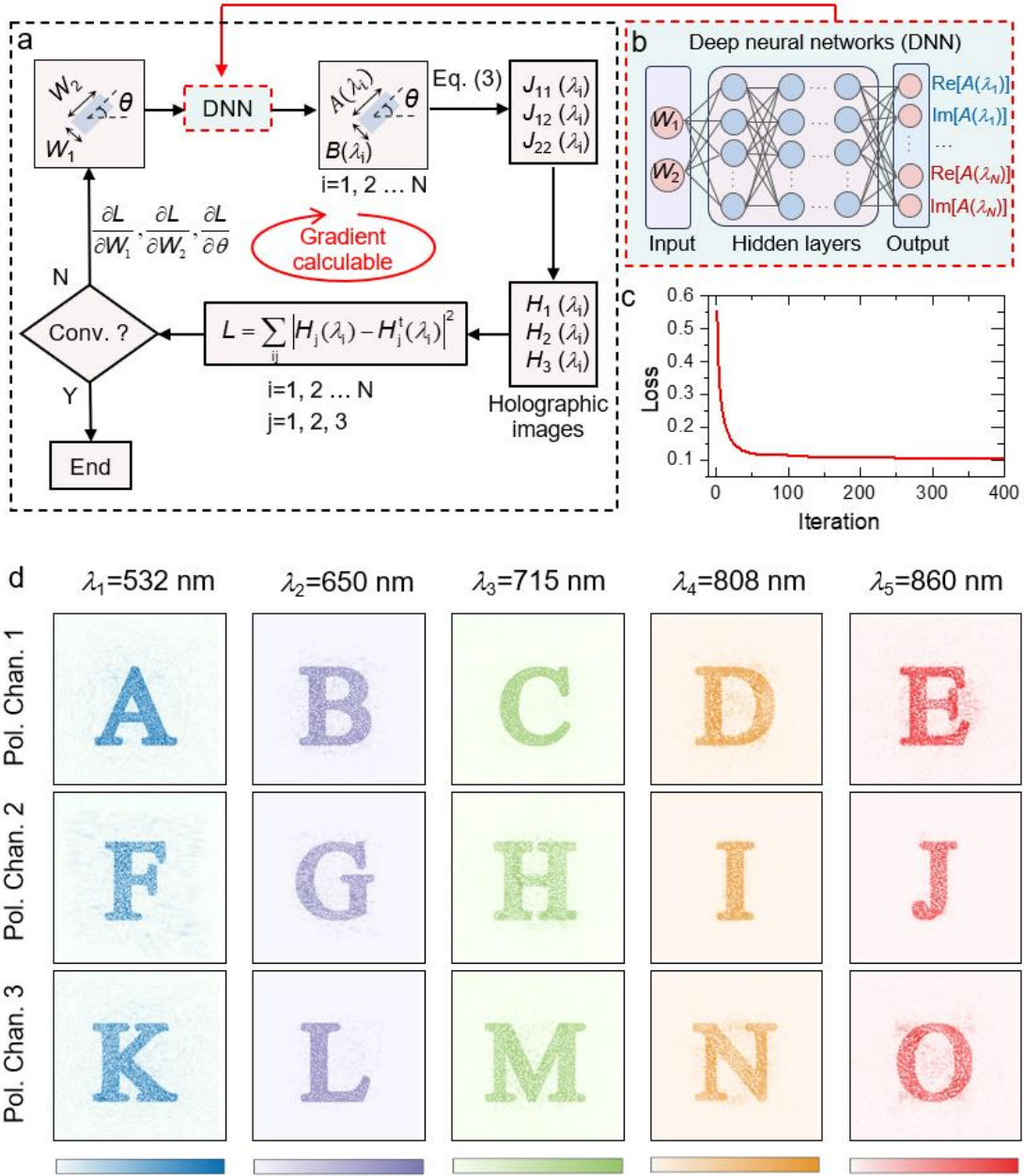

**Fig. 2. Gradient-based optimization algorithm for multiplexing limits of polarizations and wavelengths. a**, Schematic of the gradient-base optimization algorithm. The optimized parameters are the three geometric parameters of nanoblocks: $W_1$, $W_2$ and $\theta$. The loss $L$ is defined as the mean squared error between the calculated holographic images and the designed targets. The gradient of the loss $L$ with respect to the three geometric parameters facilitates the optimization algorithm. The metasurface has dimensions of 800×800 pixels, with a period of 250 nm. The holographic images are designed at 280

μm above the metasurface surface in the air. **b**, A DNN for establishing the relationship between optical coefficient *A* and geometric parameters. The DNN is a fully connected linear network with the input layer having two neurons ($W_1$ and $W_2$), and the output layer has 2*N* neurons, corresponding to the real and imaginary parts of optical coefficient *A* at *N* wavelengths. **c**, Loss value as a function of the iteration number for case with five wavelengths. **d**, Calculated holographic images under different polarization and wavelength channels.

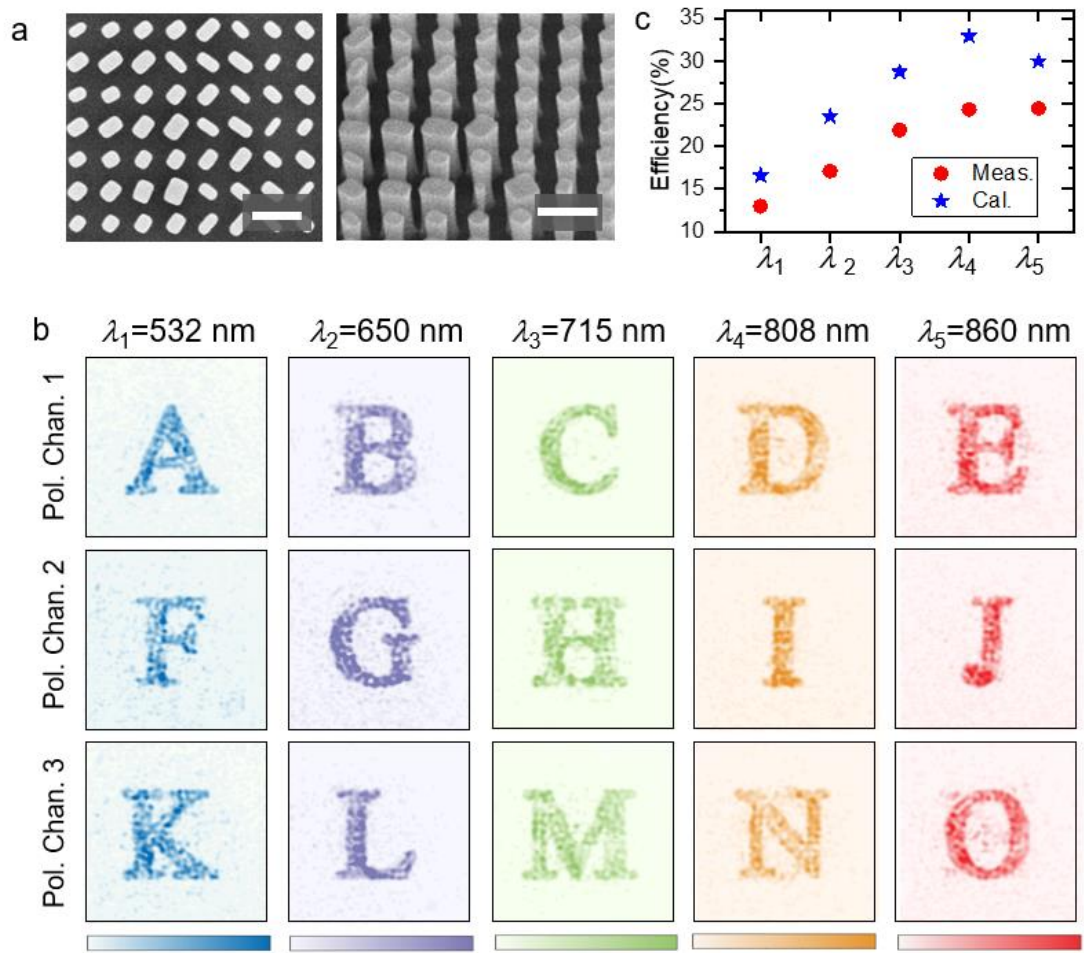

**Fig. 3. Experimental demonstration of the multiplexing limits of polarization and wavelength. a**, Scanning electron microscopy (SEM) images of the fabricated metasurface (partial view). Scale bars: 300 nm. **b**, Measured holographic images under different polarization and wavelength channels. **c**, Calculated (blue pentagons) and experimentally measured (red circles) efficiencies of the holographic
18

images under different wavelengths.

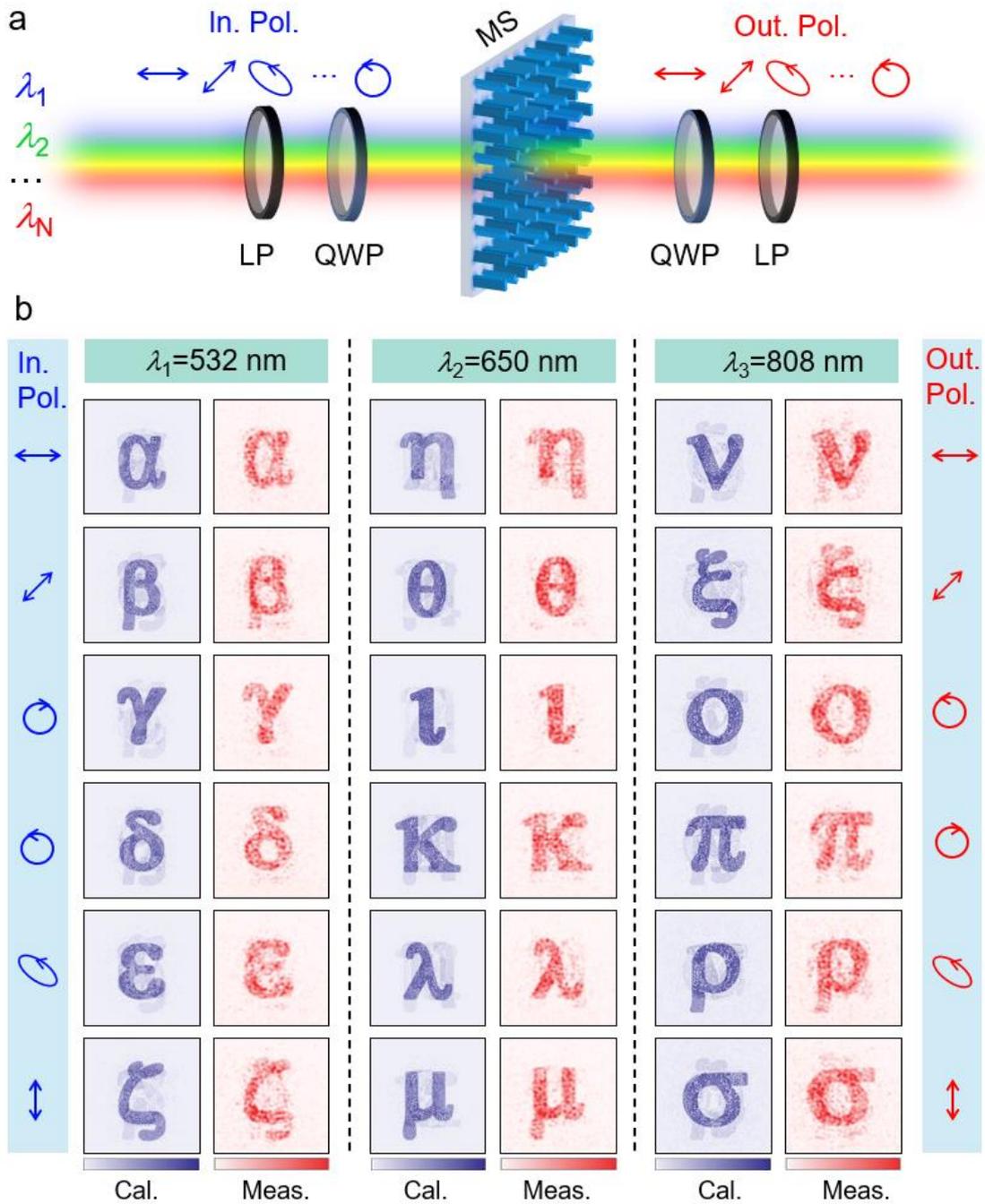

**Fig. 4. Wavelength and limit-breaking polarization multiplexing. a**, Schematic of a metasurface with wavelength and limit-breaking polarization multiplexing. The designed metasurface has independent functions under various wavelengths and more than three different combinations of input and output polarizations. The input and output polarizations are controlled by a pair of linear polarizers (LP) and quarter waveplates (QWP). **b**, Calculated and measured holographic images under three different wavelengths and six combinations of input and output polarizations.

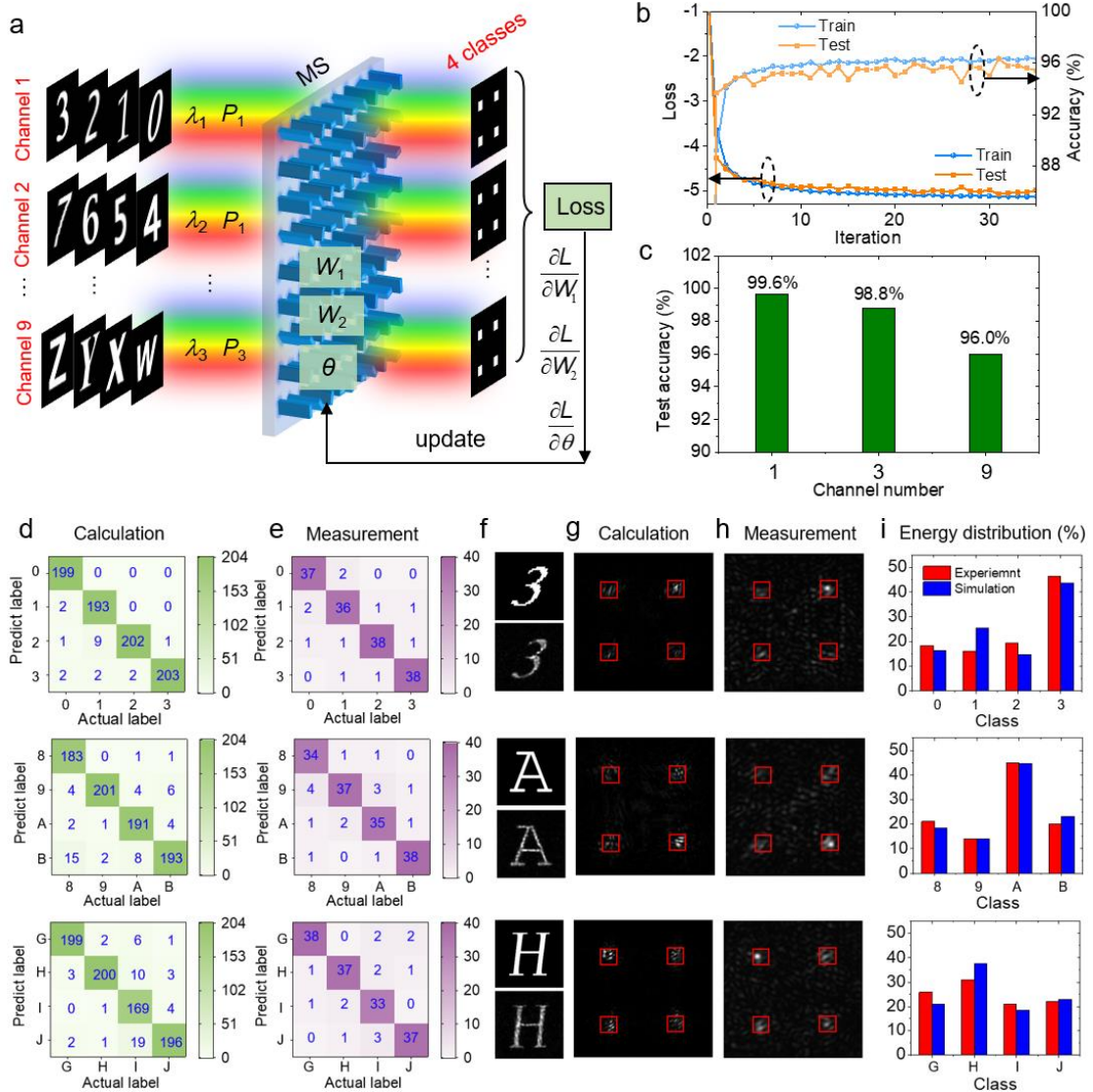

**Fig. 5. Image recognition by a single-layer metasurface with nine channels of polarization and wavelength multiplexing. a**, Schematic of a metasurface operating for nine-channel image recognition. The nine channels are the combination of three polarizations ($P_i$) and three wavelengths ($\lambda_i$) at 600 nm, 720 nm, and 840 nm. In each channel, four classes of images are classified. The metasurface contains

300×300 large pixels, with each large pixel consisting of a 4×4 array of repeated nanorods (1 μm×1 μm). The input image is placed 2 mm above the metasurface in the air, and the holographic images are detected 2 mm below the metasurface in the substrate. **b**, Loss and recognition accuracy versus iteration for both training and validation datasets. **c**, Recognition accuracy of the validation dataset for one ($J_{11}$ polarization channel at 840 nm wavelength), three (three polarization channels at 840 nm wavelength), and nine channels. **d-e**, Confusion matrices of validation datasets for calculation (d) and measurement (e) for three selected channels. **f**, Comparison between the designed input images (top) and the generated images (bottom) by an SLM in the experiment. **g-h**, Optical intensity distributions at the holographic image plane obtained in calculation (g) and measurement (h). **i**, Energy distributions of the four classes in the target regions for the experimental measurement.